\documentclass[article]{elsarticle}
\usepackage{lineno,hyperref}
\modulolinenumbers[5]
\journal{Elsevier}
\usepackage{dcolumn}   
\usepackage{bm}        
\usepackage{amssymb}   
\usepackage{lipsum}
\usepackage{amsmath,amssymb,amsfonts,graphics,graphicx,dcolumn,bm,enumerate}
\usepackage{comment,natbib,appendix}
\usepackage[ddmmyyyy,hhmmss]{datetime}
\usepackage{multirow,color}
\usepackage{adjustbox}
\usepackage{float}
\usepackage{amsthm}
\usepackage{natbib}
\usepackage{hyperref}
\usepackage{graphicx}
\usepackage{epstopdf}

\usepackage{longtable}
\usepackage{pdfpages}
\usepackage{pbox}
\usepackage{makecell}

\begin{document}
\begin{frontmatter}
\title{Identifying the global terror hubs and vulnerable motifs using complex network dynamics}
\author{Syed Shariq Husain}
\author{Kiran Sharma}
\author{Vishwas Kukreti}
\author{Anirban Chakraborti}
\address{School of Computational and Integrative Sciences,\\ Jawaharlal Nehru University, New Delhi-110067, India}
\cortext[mycorrespondingauthor]{Corresponding author: A. Chakraborti (anirban@jnu.ac.in)}

\begin{abstract}
Terrorism instills fear in the minds of people and takes away the freedom of individuals to act as they will. Terrorism has turned out to be an international menace today. Here, we study the terrorist attack incidents which occurred in the last half-century across the globe from the open source, Global Terrorism Database, and develop a view on their spatio-temporal dynamics. We construct a complex network of global terrorism and study its growth dynamics, along with the statistical properties of the anti-social network, which are quite intriguing. Normally, each nation pursues its own vision of international security based upon its mandate and particular notions of politics and its policies to counter the threat of terrorism that could naturally include the use of tactical measures and strategic negotiations, or even physical power. We study the network resilience against targeted attacks and random failures, which could guide the counter-terrorist outfits in designing strategies to fight terrorism. We then use a disparity filter method to isolate backbone of the network, and identify the terror hubs and vulnerable motifs of global terrorism. We also examine evolution of the hubs and motifs in a few exemplary cases like Afghanistan, Colombia, India, Israel, Pakistan and the United Kingdom. The dynamics of the terror hubs and the vulnerable motifs that we discover in the network backbone turn out to be very significant, and may provide deep insight on their formations and spreading, and thereby help in contending terrorism or framing public policies that can check their spread.
\end{abstract}



\end{frontmatter}

\section{Introduction}

Humans are social animals and since the early days of evolution, they have preferred to form and stay together in groups. These groups have evolved from simple settlements to huge nations; defined by multiple causes like language, common heritage, geographical boundaries, and even ideology. The human cooperation has been a motivating force behind the rapid progress of man \cite{Perc_2017}. Often evolutionary efforts induced the increase in human cooperation. Various factors, like delayed self-sustenance of young ones, or fear of elimination by neighboring communes competing for similar resources which obstructed the humans to sustain their progeny became reasons for cooperation. This cooperation extended from blood relatives to totally unrelated individuals. Surprisingly, evolution has also been responsible for drawing distinctions among themselves in their bids for the survival of the fittest. This segregation \cite{Schelling_1969,Schelling_1971} can be seen in various forms of race, caste, class, religion, political ideology, etc. Thus, the human social behavior has been extremely convoluted with multiple parameters playing crucial roles. It is extremely difficult to assess the complexity of human social behavior, which has a wide range --- co-operations, bonding, conflicts, aggression, coups, wars, etc. 

Similar to conflicts, aggression and wars, which have plagued mankind from antiquity, acts of terrorism --- where a small group of individuals which are similarly motivated in fighting another social institution or organization or exercising indiscriminate violence in achieving financial, political, religious or ideological aim is hardly new.
Though there is no single definition, terrorism may be broadly defined as a conscious and deliberate attempt to incite fear among masses through violence or the threat of violence to pursue a political or ideological gain \cite{Richardson_2013,Cutter_2014}. The aim of terrorism is not limited to eliminating the target group or destruction of opponent's resources, rather it is specifically carried out to send out a psychological message to the adversary. It is meant to propagate fear among the wider general public which may encompass rival religious or ethnic group, a state government, or an entire country. 
Even though terrorism has been prevalent ever since modern political landscape has existed, past few decades have seen an exponential increase in terrorist incidents. The scope and nature of terrorist attacks have also evolved rapidly, a fact that became evident to the world on September 11, 2001, when a series of four coordinated attacks were conducted in the United States by the terrorist group al-Qaeda. Academic and social media reports show the increase in number of terrorist acts, as well as terrorist organizations. Furthermore, the stretch of the targeted locations has extended on a global scale. These realities have made it incessantly difficult for counter-terrorist organizations or governments in terminating these terrorist acts. 

Apart from solutions by social scientists, physicists have recently tried to provide mathematical models, statistical and network analyses and potential solutions to the menaces of terrorism \cite{Galam_2003,Chakrabarti_2006,Clauset_2007}, conflicts \cite{Sharma_2017_b,sehgal2018spatio} and other social phenomena \cite{Castellano_2009,Sen_2013,husain2019dynamical}. Like business conglomerates, terrorist organizations have also formed transnational ties. They are inter-connected (in a state) and have links with other terrorist organizations outside of the geographical boundaries of the target state. 

In this paper, we develop and present a network-based study \cite{Barabasi_2016,Newman_2011} of identification of terrorist hubs and their vulnerable targets. We analyze the terrorist events obtained from the Global Terrorism Database (GTD) \cite{GTDcloud,GTDcodebook}, which collected reports from the printed and digital media, from 1970 to 2016. It is assumed that the media reports are unbiased and provide fair accounts of the actual events. 
We construct a complex network of global terrorism and investigate the network characteristics of this \textit{anti-social} network. We study the resilience of the network against targeted attacks and random failures \cite{Albert_2000,Newman_2011}, which could guide the counter-terrorist outfits in designing strategies to fight terrorism. We also use a disparity filter method \cite{Serrano_2009} to isolate the backbone of this network and identify the terror hubs and vulnerable motifs of global terrorism. We then examine the evolution of the hubs and motifs in a few special cases like Afghanistan, Colombia, India, Israel, Pakistan  and the United Kingdom. The backbone evolution reflects the change of relative importance of the terror hubs and the vulnerable motifs, which is in conformity with international affairs, peace accords, etc.

\section{Data, Methodology and Results}

\subsection{Data description and filtration}
\label{method:data}
The source for this analysis is open-access Global Terrorism Database (GTD), \cite{GTDcloud,GTDcodebook} generously provided by the National Consortium for the Study of Terrorism and Responses to Terrorism (START), University of Maryland. GTD contains the database of news articles from all over the world. The data provides a detailed account of terrorist events from 1970 to 2016, except for year 1993 for which no data exists. The dataset has 170350 instances divided into 135 attributes.
For each instance/event-data, we filtered information such as source, target, date of event, organizations or groups, the location information of the event, as well as latitude, longitude data of actors and the event.
The dataset required considerable cleaning before any study could be done. The doubtful events, suggested by dataset itself, were removed. Further, attacks carried out by ``Unknown'' terrorist organization on ``Unknown'' targets were also filtered out. The events which had no spatial information in the dataset were removed, too. To maintain the network modularity at the country level, any attack which targeted international community instead of a particular nationality was also not included in the dataset. We finally had 64855 events in 46 years (no records available for year 1993).

\subsection{Network construction}
\label{method:network}

We analyze the terrorist attacks over a 46 year period (1970-2016), excluding 1993 (non-availability of data for the given period), such that the temporal granularity of our data analyses is one day. Over a period of time $\tau$, we construct the network of `connected' actors in the following way: 
Whenever a terrorist source, actor $a_1$, attacks a target, actor  $a_2$, it is recorded as an event $E_1$ at time $t \in [t_0 : t_0 + \tau ]$, an edge or directed link `connects' the source to the target by an arrow of unit weight (could be generalized by weighing the edge by impact of attack, etc.), where $t_0$ is the initial time in the entire span $\tau$. If another event $E_2$ within the same time window involves source $a_3$ and target $a_2$, then $a_3$ is connected to $a_2$ with a directed link of unit weight. Thus, sources $a_1$ and $a_3$ are both connected to the common target $a_2$. Aggregating all such events over the time window $\tau$, connected components are formed. Every time the same pair of actors are engaged, the edge mention (weight) increases by unity. Thus, the weight of an edge is the frequency of occurrence of attacks between the pair of source and target. 

\subsection{Disparity filter}
To find the backbone structure of a weighted network, we have  used an algorithm proposed by Serrano et al.~\cite{Serrano_2009}. The disparity filter algorithm extracts the network backbone by considering the relevant edges at all the scales present in the system and exploiting the local heterogeneity and local correlations among the weights. The disparity filter has a cut-off parameter $\alpha_c$, which determines the number of edges that are reduced in the original network. The filter, however, preserves the cut-off of the degree distribution, the form of the weight distribution, and the clustering coefficient. We discuss the choice of the cut-off parameter $\alpha_c$ below.

\subsection{Results}
\subsubsection{Network structure and backbone using disparity filter}
\begin{figure}[h!]
\centering
\includegraphics[width=.95\linewidth]{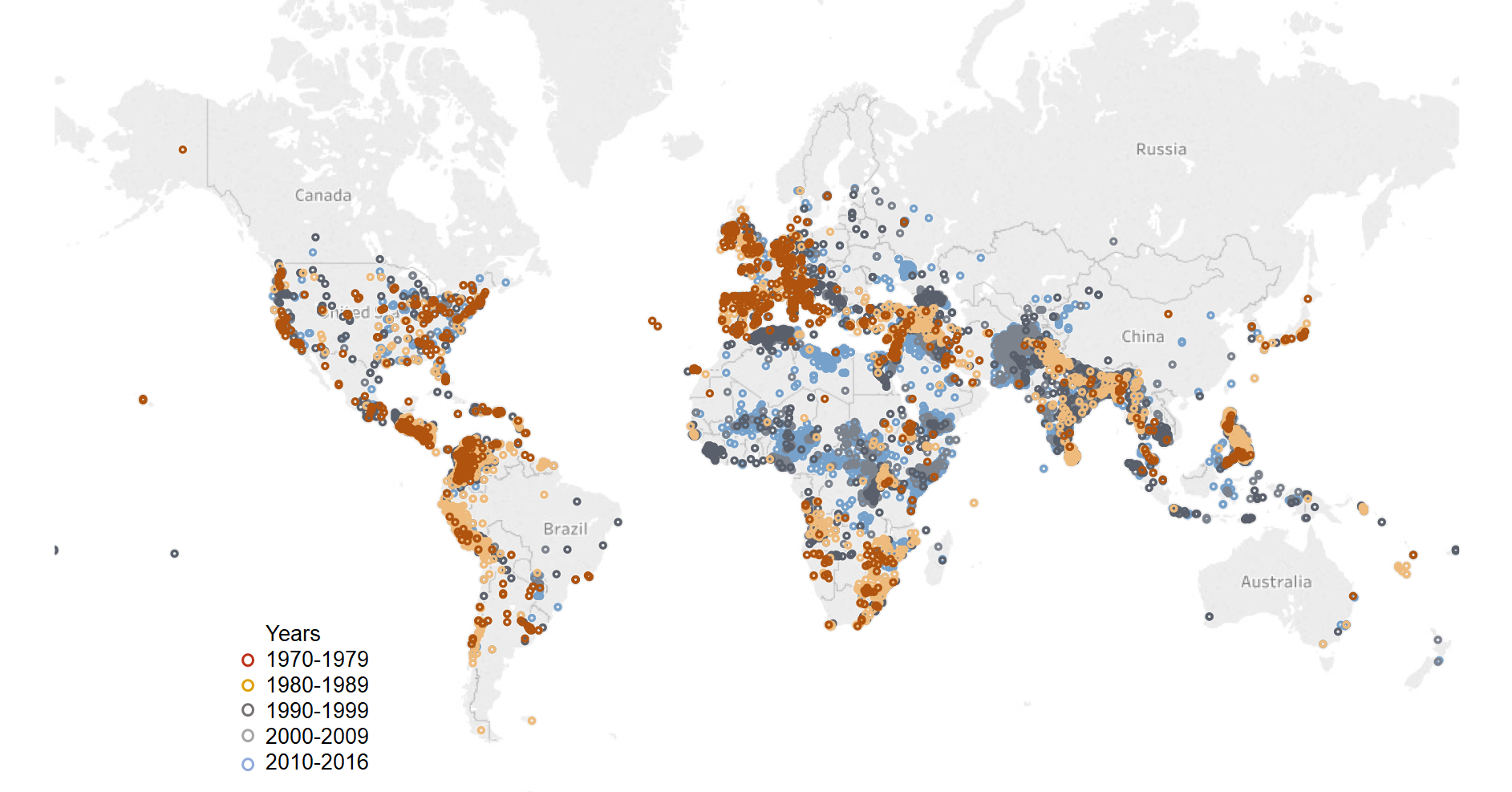}
\llap{\parbox[b]{4.5in}{\textbf{(a)}\\\rule{0ex}{2.2in}}}
\\\includegraphics[width=0.4\linewidth]{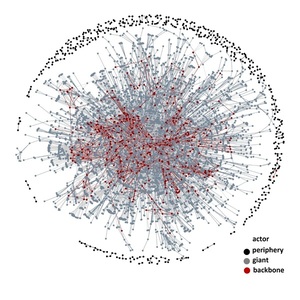}
\llap{\parbox[b]{2.1in}{\textbf{(b)}\\\rule{0ex}{1.6in}}}
\includegraphics[width=0.4\linewidth]{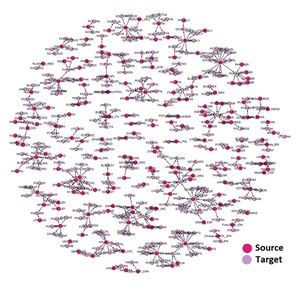}
\llap{\parbox[b]{2in}{\textbf{(c)}\\\rule{0ex}{1.6in}}}
\caption{(a) Spatio-temporal evolution of attacks from 1970-2016. The different colored dots indicate the locations of the attacks, along with the years mentioned in the legend. The map and international boundaries are generated using the proprietary software \textit{Tableau Desktop version 10.5.0}. (b) The aggregated network of terrorism for the period 1970-2016 -- Network of terrorist attacks constructed from the history of the events. Two types of actors are involved in the network: Source and target (source nodes refer to the terrorist organizations and target nodes are the victims). Each actor is a node and whenever two actors are involved in an event, a directed link is drawn from source to target. The network has a giant component (grey) in the center, surrounded by 168 peripheral isolated clusters (black), while the nodes colored in red are showing the backbone; the average degree of a node in the network is $3.718$. (c) The zoomed-in view of the backbone of the network, which has been identified using the disparity filter with $\alpha_c=0.01$ (described above). The backbone contains $8\%$ of the total nodes and $4\%$ of total edges (details summarized in Table~\ref{table:table1})}.
\label{fig:World_and_Network}
\end{figure}

Fig.~\ref{fig:World_and_Network} (a) shows the spatio-temporal distribution of the events across the globe; the dots representing the locations, colored according to different decades during which the events occurred. Evidently, past few decades have seen an exponential increase in terrorist incidents (see Supplementary Information Fig. S2, where year-wise evolution along with the associated frequencies of events are depicted).  Note that during the period 1970-1990, the maximum value stayed at 2186, while during the period of 1995-2015 this value shot from 364 to 5703, with some fluctuations in between.
Fig.~\ref{fig:World_and_Network} (b) shows a network aggregated over the period 1970-2016. Two types of actors are involved in the network: source and target (source nodes refer to the terrorist organizations and target nodes are the victims). Each actor is a node and whenever two actors are involved in an event, a directed link is drawn from source to target. The network has a giant component (colored in grey) in the center, surrounded by 168 peripheral isolated clusters (colored in black), while the nodes colored in red are showing the backbone. The zoomed-in view of the backbone structure is shown in Fig.~\ref{fig:World_and_Network} (c).
The terrorist network for the entire period consists of $5568$ nodes, $64855$ edge mentions, and $10379$ unique edges, with the giant component consists of $5148$ nodes. Thus, the giant component encompasses more than $92\%$ of the network nodes, implying that terrorism is closely linked across the globe and the backbone structure encompasses $8\%$ of the total nodes of the network with disparity filter~\cite{Serrano_2009} cut-off $\alpha_c=0.01$ during the period 1970-2016.

\begin{figure}[ht!] 
\centering
\includegraphics[width=0.45\linewidth]{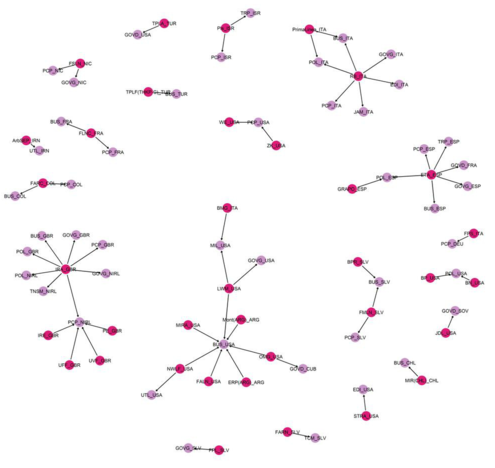}
\llap{\parbox[b]{2.4in}{\textbf{(a) 1970-80}\\\rule{0ex}{2.in}}}
\includegraphics[width=0.45\linewidth]{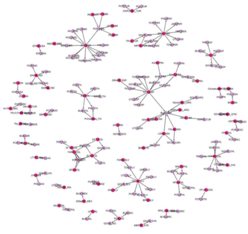}
\llap{\parbox[b]{2.4in}{\textbf{(b) 1970-90}\\\rule{0ex}{2in}}}
\includegraphics[width=0.45\linewidth]{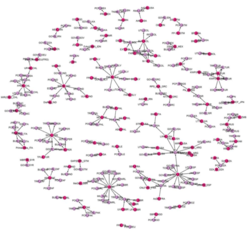}
\llap{\parbox[b]{2.4in}{\textbf{(c) 1970-2000}\\\rule{0ex}{2in}}}
\includegraphics[width=0.45\linewidth]{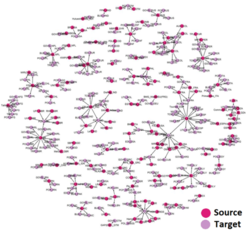}
\llap{\parbox[b]{2.4in}{\textbf{(d) 1970-2010}\\\rule{0ex}{2in}}}
\caption{Decade-wise evolution of the network backbone from 1970-2010 (a-d) (for 1970-2016 shown in Fig.~\ref{fig:World_and_Network}). The zoomed-in views of the backbones of the growing network, which have been identified using the disparity filter with $\alpha_c=0.01$. The characteristics of the growing backbone structure for the different decades are summarized in Table\ref{table:table1}.}
\label{fig:backbone_evolution}
\end{figure}

The decade-wise evolution of the aggregate network, its giant component (in grey) as well as the backbone (in red) are shown in Fig.~\ref{fig:giant_comp} of Supplementary Information. Here, we focus more on the evolution of the backbone. Fig.~\ref{fig:backbone_evolution} (a-d) show the decade-wise evolution of the network backbone from $1970$ to $2010$, identified using the disparity filter with $\alpha_c=0.01$. The comparison of backbone characteristics (as percentage of the total weights $\%W_T$ of the network, total nodes $\%N_T$ in the network and total unique edges $\%E_T$ in the network) with the change in disparity filter cut-off $\alpha_c$, for the different evolving backbones for the period 1970-1980, 1970-1990, 1970-2000, 1970-2010, and 1970-2016, is summarized in Table \ref{table:table2}. We have studied other choices of $\alpha_c$ (see Fig. S5 in Supplementary Information) as well, to check the granularity and details of motifs. In this paper, for filtering the backbone of the network, we choose $\alpha_c$ at 0.01, which captures the motifs at the country level efficiently and provides fairly accurate insights in terms of the participation of the actors (sources and targets). 
 
\begin{table}
\centering
\caption{Characteristics of the growing backbone structures (with $\alpha_c=0.01$) in  Fig. \ref{fig:backbone_evolution}, for the different decades.}
\label{table:table1}
\vspace*{2mm}
\begin{tabular}{|c|c|c|c|c|c|}
\hline
Year      & Nodes (\%) & Edges (\%) & Edge mentions (\%) & \begin{tabular}[c]{@{}c@{}}Number of \\ clusters\end{tabular} & \begin{tabular}[c]{@{}c@{}}Average number\\ of neighbors\end{tabular} \\ \hline
1970-1980 & 82 (16)   & 63 (3)     & 2490 (38)          & 20                                                            & 1.537                                                                 \\ \hline
1970-1990 & 179 (17)  & 156 (4)    & 12440 (56)         & 35                                                            & 1.743                                                                 \\ \hline
1970-2000 & 265 (7)   & 230 (4)    & 17730 (57)         & 51                                                            & 1.736                                                                 \\ \hline
1970-2010 & 341 (7)   & 308 (4)    & 23707 (57)         & 60                                                            & 1.806                                                                 \\ \hline
1970-2016 & 470 (8)   & 427 (4)    & 40467 (62)         & 79                                                            & 1.817                                                                 \\ \hline
\end{tabular}
\end{table}

\begin{table}[]
\centering
\caption{Comparison of backbone characteristics (as percentage figures of the total weights $\%W_T$  of the network, total nodes $\%N_T$ in the network and total unique edges $\%E_T$ in the network) with the change in disparity filter cut-off $\alpha_c$, for the different evolving backbones.}
\label{table:table2}
\vspace*{2mm}
\begin{tabular}{|l|l|l|l|l|}
\hline
Year              & $\alpha_c$ & $\%W_T$ & $\%N_T$ & $\%E_T$ \\ 
 \hline
\multirow{5}{*}{1970-1980}                       & 0.05 & 48   & 9    & 6    \\ \cline{2-5} 
                                                 & 0.04 & 45   & 9    & 5    \\ \cline{2-5} 
                                                 & 0.03 & 43   & 8    & 4    \\ \cline{2-5} 
                                                 & 0.02 & 40   & 7    & 4    \\ \cline{2-5} 
                                                 & 0.01 & 38   & 6    & 3    \\ \hline
Year              & $\alpha_c$ & $\%W_T$ & $\%N_T$ & $\%E_T$ \\ 
 \hline
\multirow{5}{*}{1970-1990}                       & 0.05 & 65   & 11   & 7    \\ \cline{2-5} 
                                                 & 0.04 & 64   & 10   & 6    \\ \cline{2-5} 
                                                 & 0.03 & 62   & 10   & 5    \\ \cline{2-5} 
                                                 & 0.02 & 60   & 8    & 5    \\ \cline{2-5} 
                                                 & 0.01 & 56   & 7    & 4    \\ \hline
Year              & $\alpha_c$ & $\%W_T$ & $\%N_T$ & $\%E_T$ \\ 
 \hline
\multicolumn{1}{|c|}{\multirow{5}{*}{1970-2000}} & 0.05 & 65   & 11   & 6    \\ \cline{2-5} 
\multicolumn{1}{|c|}{}                           & 0.04 & 64   & 10   & 6    \\ \cline{2-5} 
\multicolumn{1}{|c|}{}                           & 0.03 & 63   & 9    & 5    \\ \cline{2-5} 
\multicolumn{1}{|c|}{}                           & 0.02 & 61   & 8    & 5    \\ \cline{2-5} 
\multicolumn{1}{|c|}{}                           & 0.01 & 57   & 7    & 4    \\ \hline
Year              & $\alpha_c$ & $\%W_T$ & $\%N_T$ & $\%E_T$ \\ 
\hline
\multirow{5}{*}{1970-2010}                       & 0.05 & 65   & 12   & 7    \\ \cline{2-5} 
                                                 & 0.04 & 64   & 11   & 6    \\ \cline{2-5} 
                                                 & 0.03 & 63   & 10   & 5    \\ \cline{2-5} 
                                                 & 0.02 & 60   & 9    & 5    \\ \cline{2-5} 
                                                 & 0.01 & 57   & 7    & 4    \\ \hline
Year              & $\alpha_c$ & $\%W_T$ & $\%N_T$ & $\%E_T$ \\ 
\hline
\multirow{5}{*}{1970-2016}                       & 0.05 & 70   & 13   & 7    \\ \cline{2-5} 
                                                 & 0.04 & 68   & 12   & 6    \\ \cline{2-5} 
                                                 & 0.03 & 67   & 11   & 6    \\ \cline{2-5} 
                                                 & 0.02 & 65   & 10   & 5    \\ \cline{2-5} 
                                                 & 0.01 & 62   & 8    & 4    \\ \hline
\end{tabular}

\end{table}

\subsubsection{Network statistical properties}
As mentioned earlier, we construct the aggregate network over a given period of time, for the entire 46 years period (1970-2016) as well as for decade-wise period 1970-1980, 1970-1990, 1990-2000, and 1970-2010. Each event is visualized as a link between the source and target involved, which are the nodes, thus creating a network of actors connected by events. The details of the construction is given in the Methods section~\ref{method:network}. A typical network is shown in Fig.~\ref{fig:World_and_Network} (a) for data aggregated over the year 1970-2016.

We extracted the number of edge mentions $w$ of a unique pair of actors (source-target), as well as the number of unique actors $k$, one actor is involved with. In terms of the network theory,  $w$ is the link weight and $k$ the degree of a node (out-degree for sources and in-degree for targets). While $k$ measures the importance, activity or visibility of a single actor, $w$ measures the frequency of involvement of an actor pair (source-target) in the terrorist events. Since this is a directed network, the nodes have distinct out-degree and in-degree distributions. These quantities are measured decade-wise for the 46 year period (1970-2016) as well as through the whole period (aggregate over 46 years).

\begin{figure}[h!] 
\includegraphics[width=0.46\linewidth]{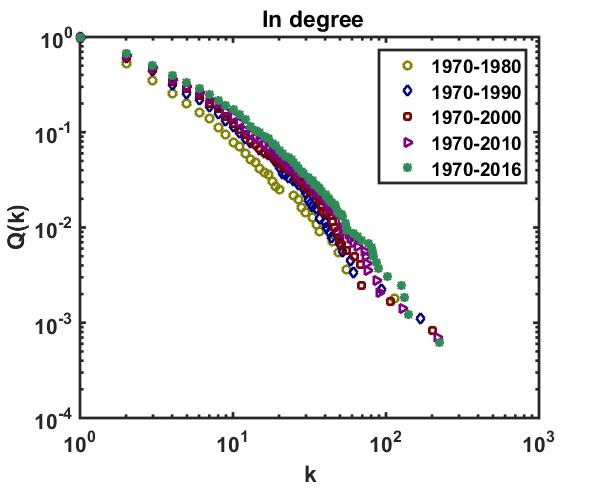}
\llap{\parbox[b]{2.3in}{\textbf{(a)}\\\rule{0ex}{1.7in}}}
\includegraphics[width=0.46\linewidth]{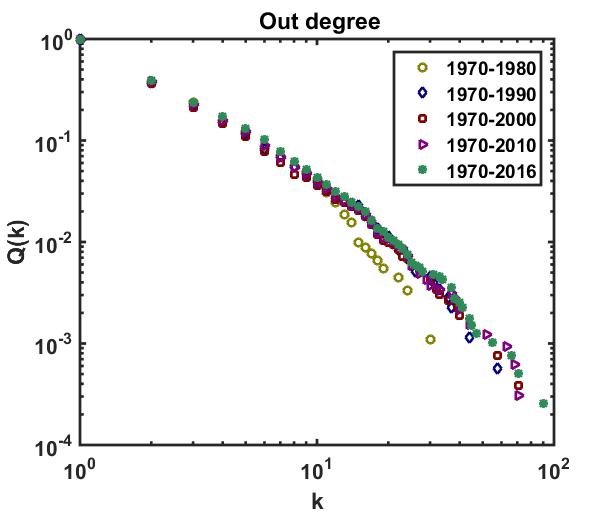}
\llap{\parbox[b]{2.3in}{\textbf{(b)}\\\rule{0ex}{1.7in}}}\\
\includegraphics[width=0.46\linewidth]{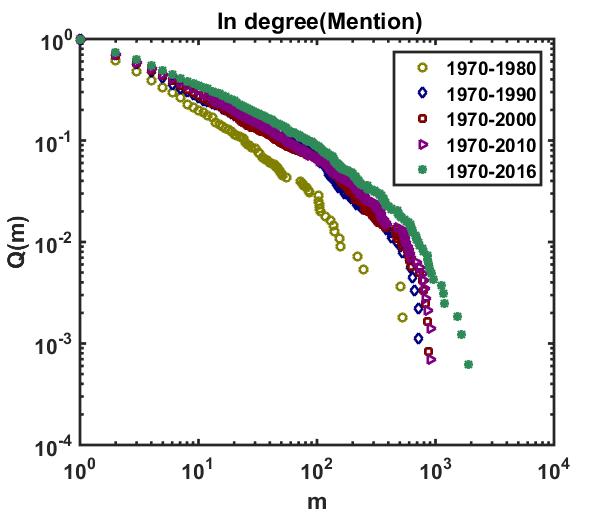}
\llap{\parbox[b]{2.3in}{\textbf{(c)}\\\rule{0ex}{1.7in}}}
\includegraphics[width=0.46\linewidth]{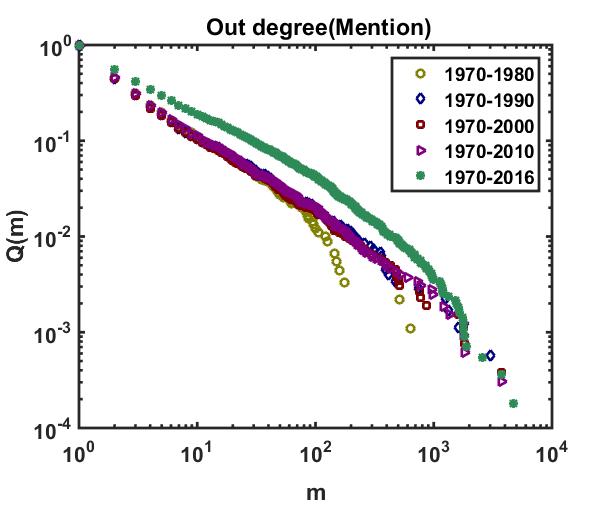}
\llap{\parbox[b]{2.3in}{\textbf{(d)}\\\rule{0ex}{1.7in}}}\\
\includegraphics[width=0.46\linewidth]{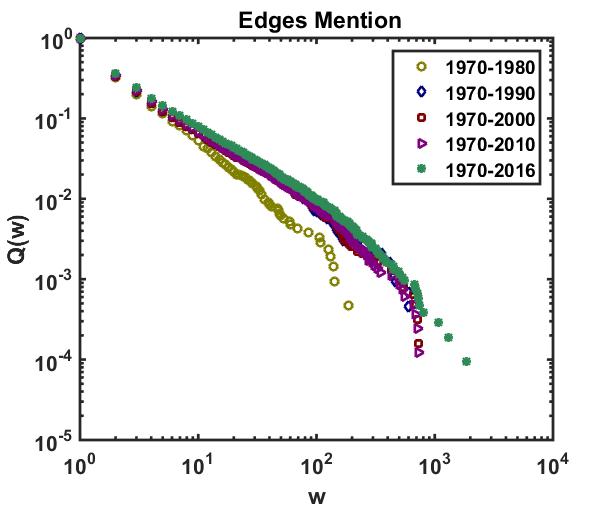}
\llap{\parbox[b]{2.3in}{\textbf{(e)}\\\rule{0ex}{1.7in}}}
\includegraphics[width=0.46\linewidth]{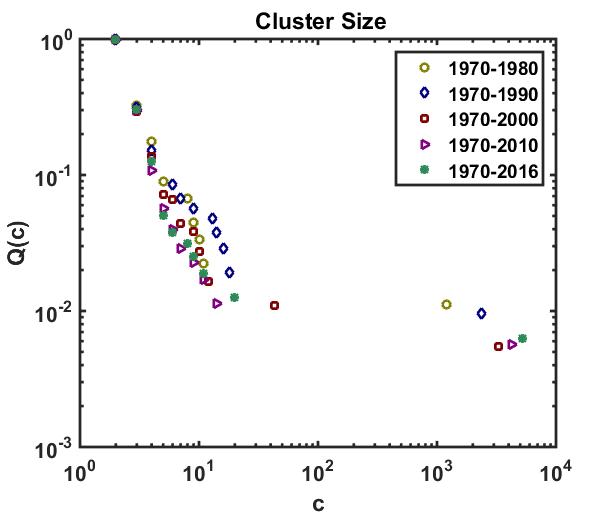}
\llap{\parbox[b]{2.3in}{\textbf{(f)}\\\rule{0ex}{1.7in}}}
\caption{Structural properties of the network decade-wise: 1970-1980 (light green circle), 1970-1990 (blue diamond), 1970-2000 (red square). 1970-2010 (magenta triangle) and 1970-2016 (green star). (a and b) Plot of the cumulative probability (CCDF) $Q(k)$ that an actor is connected to $k $ others or more (in-degree and out-degree). 
(c and d) Plot of the cumulative probability (CCDF) $Q(m)$ that an actor is mentioned at least $m$ times (in-degree and out-degree). (e) Plot of the cumulative probability $Q(w)$ that an actor pair is mentioned at least $w$ times. (f) Plot of the cumulative probability (CCDF) for $Q(c)$ that there is a cluster of size larger than $c$. The size of the largest clusters (seen as outliers) are very large compared to the rest.}
\label{fig:dynamics}
\end{figure}

We computed the complementary cumulative probability distribution (CCDF) for different quantities for the aggregated network in 1970-1980, 1970-1990, 1970-2000, 1970-2010 and 1970-2016 (refer to Fig.~\ref{fig:dynamics} a-f).
The complementary cumulative probability density functions for degree  $Q(k)$ (out- and in-degrees) and the edge-mentions $Q(w)$, which appear to be broad distributions (stretched-exponentials) (see Fig.~\ref{fig:dynamics}).
The CCDF's $Q(s)$ of the cluster size $s$ for the growing networks are shown in Fig.~\ref{fig:dynamics} (f); evidently, the outliers correspond to the sizes of the growing giant cluster (as percentage of the total number of nodes in the network): 1210 ($83\%$) (1970-80), 2346 ($88\%$) (1970-90), 3313 ($86\%$) (1970-2000), 4220 ($90\%$) (1970-2010) and 5148 ($92\%$) (1970-2016).

The above results quantitatively characterize the heterogeneity in the activities of the different actors (sources and targets)-- while most actors (or actor-pairs) are relatively less active, a few are very active (see table ~\ref{table:table3} for the list of top-50 actor pairs).
The broad degree distributions show little change with time. Notably, the average clustering coefficient (for directed network \citep{fagiolo2007clustering,malliaros2013clustering})  of the nodes in the backbone is \textit{zero}, contrary to most social networks of friendships, collaborations, etc., where typically the average clustering coefficient is high~\cite{Barabasi_2016,Newman_2011}.  This implies the absence of loops or cyclicity in the network.

%
\subsubsection{Tolerance to attack and failure.}

\begin{figure}[h!]
\includegraphics[width=0.95\linewidth]{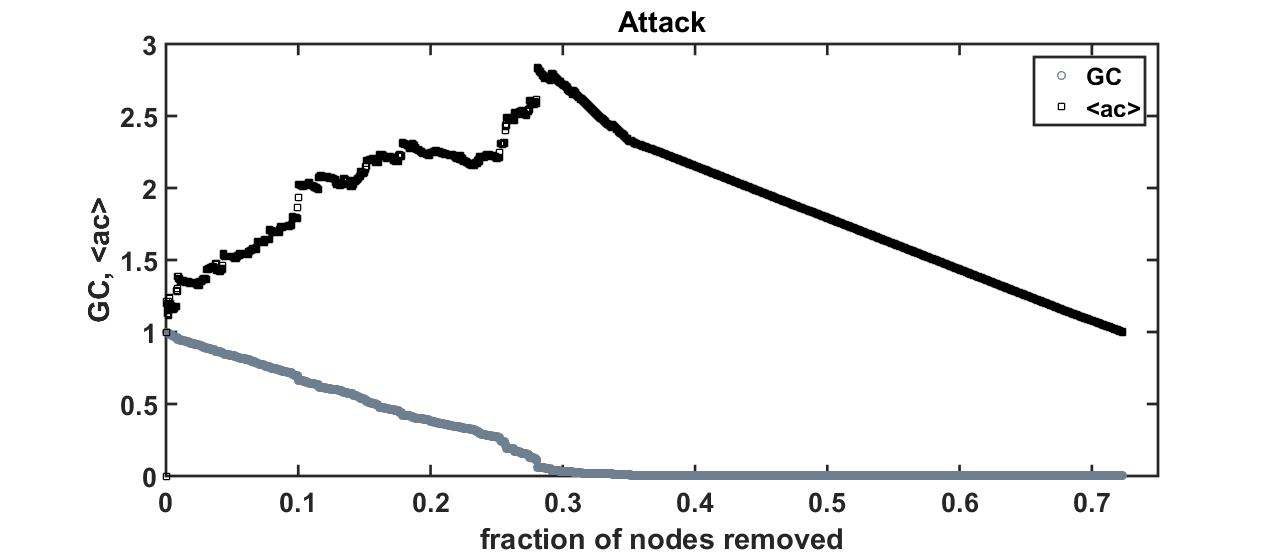}
\llap{\parbox[b]{4.7in}{\textbf{(a)}\\\rule{0ex}{1.8in}}}
\includegraphics[width=0.95\linewidth]{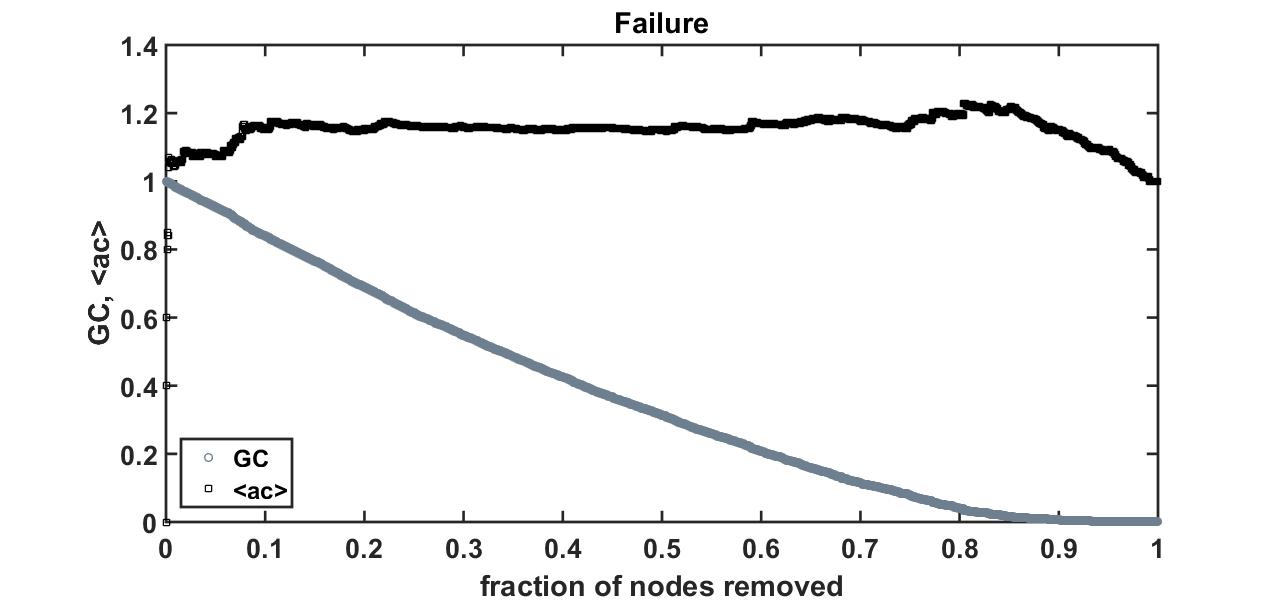}
\llap{\parbox[b]{4.5in}{\textbf{(b)}\\\rule{0ex}{2in}}}
\caption{(a) The structure of the network (directed) under targeted attack: Terrorist source nodes are removed in the sequence of their out-degree starting from the highest out-degree. The plots shows the behavior of the giant component $GC$ (fraction of nodes in the largest connected component) and the average number of nodes in the isolated clusters other than the giant component $\langle ac \rangle$, with increasing fraction of removed nodes.  (b) The structure of the network (directed) under random failure: Nodes having out-degree are removed randomly. The network and the giant component are destroyed faster by targeted nodes removal (attack), compared to the random node removal (failure): $GC$ becomes zero after about $33\%$ of the sources are removed through the former method, and about $87\%$ of the sources are removed through the latter. The results are shown for the network aggregated over $1970-2016$. }
\label{fig:resilience}
\end{figure}
We studied the network resilience-- how the network breaks down under attack, in order to stop terrorism activities to happen \cite{Albert_2000}. The largest connected component of the network (i.e., the giant component) is subjected to targeted attack by removal of the most connected nodes (in terms of the source out-degree, which corresponds to a terrorist organization, etc.). As the network is directed, so we start removing the source node with the highest out-degree, followed by the next highest out-degree and so on. This results in rapid fragmentation or destruction of the network by removing all the source nodes in comparison with random node removal. We compute the fraction of nodes present in the largest connected component  (\textit{GC}), which is observed to decrease very quickly, and the average number of nodes in the isolated clusters other than the giant component $\langle ac \rangle$, with increasing fraction of removed nodes. The network and the giant component are destroyed faster by targeted nodes removal (attack), compared to the random node removal (failure), $GC$ becomes zero after about $33\%$ of the sources are removed through the former method, and about $87\%$ of the sources are removed through the latter, as shown in Fig.~\ref{fig:resilience}. The results are very similar to those of the studies by Sharma et al. \citep{Sharma_2017_b}, but we reiterate that in this case we have a directed network and we remove only the source nodes (terrorists).

\subsubsection{Motifs and hubs}
\begin{figure}[h!]
 \includegraphics[width=0.97\linewidth]{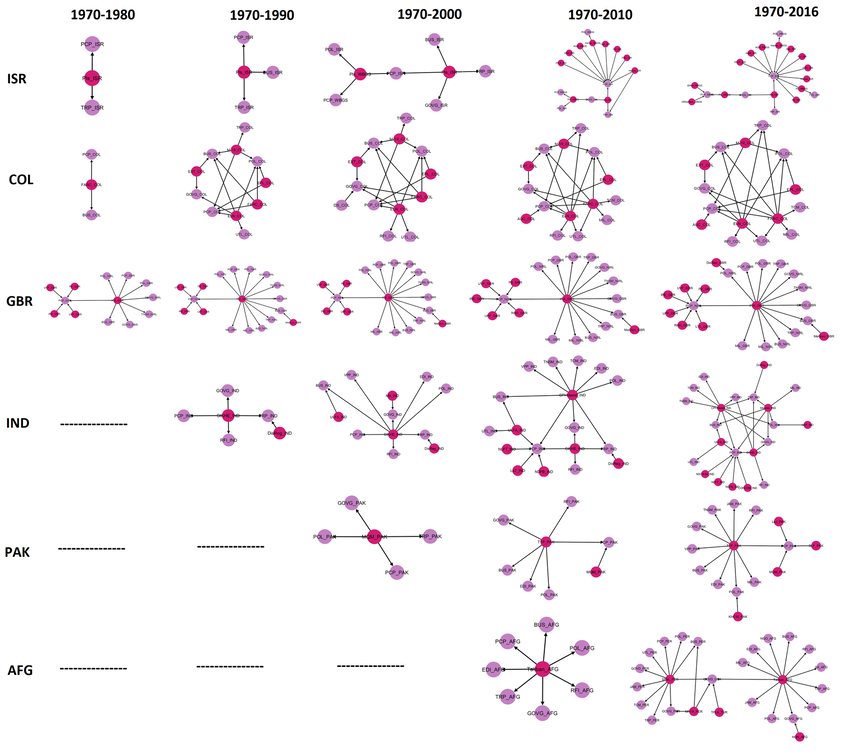}
\caption{Cumulative growth of the terror hubs and vulnerable targets in backbones of the countries: Israel (ISR), Columbia (COL), the United Kingdom (GBR), India (IND), Pakistan (PAK) and Afghanistan (AFG) from Fig. \ref{fig:World_and_Network}.}
\label{fig:motifs}
\end{figure}

Using the disparity filter method, we isolated the backbone of network for different time periods and identified the terror hubs and vulnerable motifs of global terrorism.
As obvious, only the terror hubs and vulnerable motifs that are very frequently engaged do appear in the backbone. We show in Fig.~\ref{fig:motifs}, the evolution of  hubs and motifs in a few exemplary cases like Afghanistan (AFG), Colombia (COL), India (IND), Israel (ISR), Pakistan (PAK) and the United Kingdom (GBR). The very fact that the backbone structure evolves indicates that often some terrorist organizations gain more prominence than others.
Examining these hubs and motifs, we observe that the \textit{star-structure} occurs quite frequently in the backbone: One source attacking many targets, or one target being attacked by many sources. The backbone structure of ISR grows from a simple structure of 1 source and 2 targets (average degree 1.33) in 1970-1980, to an intricate structure of 14 sources and 6 targets (average degree 1.90) in 1970-2016. Further analysis reveals that before 2000 the Israel-Palestinian conflict was limited to the aggression between the two states. Later, Palestinian organizations like Hamas, Popular Front for the Liberation of Palestine, al-Aqsa Martyr's Brigade, etc., assumed prominence in the conflicts against the state of Israel.
The backbone structure of GBR remains fairly the same; the average degree grows from 1.85 (1970-1980) to 1.91 (1970-2016); the Irish Republican Army is the main terrorist hub, while the private citizens and property is the most vulnerable target for the entire duration. In COL, the backbone structure grows from a simple 3 node (1 source and 2 targets) in 1970-1980 to a clustered 11 nodes (5 sources and 6 targets) in 1970-1990; the average degree jumps from 1.33 to 2.91. Then it grows steadily to a closely knit structure of 15 nodes (6 sources and 9 targets) at the end of 2016. We observe that till 2000 there is no node for the military, but it appears later as a target. This relates to the fact that military got involved by the state for the eradication of terrorists after the presidential change in 2002.

\begin{figure}[h!] 

\includegraphics[width=0.95\linewidth]{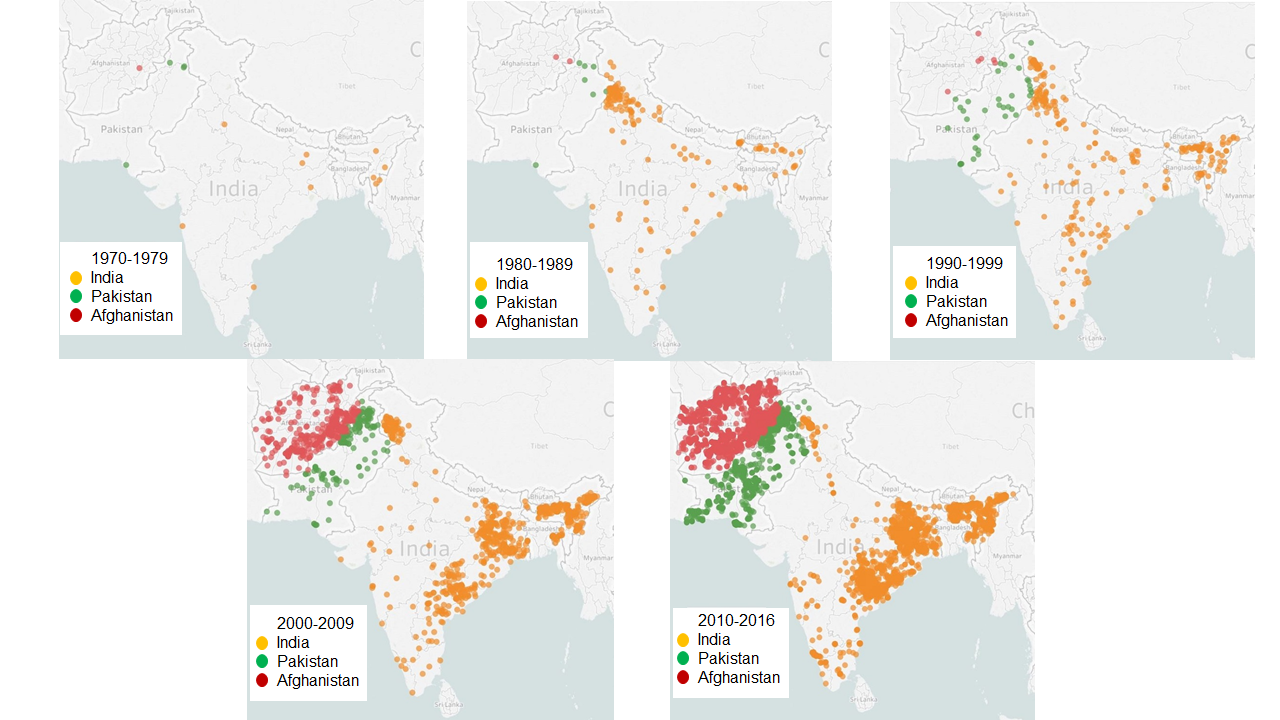}
\llap{\parbox[b]{4.5in}{\textbf{(a)}\\\rule{0ex}{2.6in}}}
\llap{\parbox[b]{3.1in}{\textbf{(b)}\\\rule{0ex}{2.6in}}}
\llap{\parbox[b]{1.7in}{\textbf{(c)}\\\rule{0ex}{2.6in}}}
\llap{\parbox[b]{4.1in}{\textbf{(d)}\\\rule{0ex}{1in}}}
\llap{\parbox[b]{2.6in}{\textbf{(e)}\\\rule{0ex}{1in}}}

\caption{Spatio-temporal evolution of Indian sub-continent. Event locations decade-wise (a-e), for the countries: Afghanistan (AFG), India (IND) and Pakistan (PAK). The maps and international boundaries are generated using the proprietary software \textit{Tableau Desktop version 10.5.0}. The results are in conformity with the backbone structures of the actor pairs of countries shown in Table 3 and Fig. \ref{fig:backbone_evolution} respectively.}
\label{fig:IND_PAK_AFG}
\end{figure}

Fig.~\ref{fig:IND_PAK_AFG} shows that in Indian sub-continent, AFG, IND, and PAK are among the nations which have suffered highest terrorist attacks in the recent decades (see Supplementary Information Fig. S2).
Interestingly, AFG does not appear in the backbone till 2000, and government (Diplomatic) GOVD\_USA is a common target which links two countries AFG and PER. This is typically the case in many other empirical networks, where there exists a node connecting two modules or communities \cite{Onnela_2007}. In AFG and PER, we again see the appearance of star structures, as in GBR. In IND, the rise (and fall) of the attacks by Sikh extremists in the state of Punjab (1980-2000), Separatists India in the state of Jammu and Kashmir (1990-2016), and Maoists in the states of Andhra Pradesh, Bihar, Chattisgarh, Jharkhand, Odisha and West Bengal (2000-2016), are noteworthy.  
The backbone structure of PAK significantly exposes the change in the prominence of terrorist organizations in various decades. With Muttahida Qami Movement having prominence in the early decades, it is replaced by Tehrik-i-Taliban Pakistan, formed in 2007 in the 2010 decade. Also, Khorasan Chapter of the Islamic State, Lashkar-e-Jhangvi, Baloch Liberation Front  have increased their attacks in the current decade, rightly captured in the backbone.

In addition, we have observed in Fig. \ref{fig:backbone_evolution} that El Salvador (SLV) appeared in 1970-1980 and 1970-1990 backbones, but did not appear thereafter; this conforms to the fact that Chapultepec peace accords were signed in 1992 and the terrorist attacks reduced.



\section{Discussion}
We have examined the spatio-temporal dynamics of the terrorist events across the globe, using the Global Terrorism Database (GTD) \cite{GTDcloud,GTDcodebook} from 1970 to 2016. We developed the view of a complex network of global terrorism and studied its growth dynamics. The network always had a giant component, which was in the range of $83\%$ to $92\%$ of the total number of network nodes. The statistical properties of the network were found to be quite robust. The CCDF's for the (in- and out-) degrees ($k$) and edge mentions ($w$) seem to be broad (stretched-exponentials).
The network resilience results yielded that the giant component disappeared after about $33\%$ of the hubs (in descending order of magnitude) were removed; in the case of random removal of sources, the giant component disappears much slower-- only after $87\%$ of the sources were removed.

We isolated the backbone of the terrorist network using the disparity filter method, and identified the terror hubs and vulnerable motifs of global terrorism. 
We have chosen $\alpha_c=0.01$ such that it enables us to follow the country-wise evolution of the terror hubs and vulnerable motifs that appear in the backbone structures, and we extracted the backbones of the networks for the different periods of evolution. Fig. \ref{fig:World_and_Network} (c) shows the backbone for the network with the aggregated data from  $1970$ to $2016$, consisting of $470$ nodes ($8\%$ of the total network), $427$ edges ($4\%$ of the total network) and $40467$ edge mentions ($62\%$ of the total network). The backbone nodes primarily lie in the giant component, with few in the periphery.
The complete lists of 190 unique sources and 280 unique targets (total $470$ nodes) are given in Tables S2-S3  in the Supplementary Information. Fig. \ref{fig:backbone_evolution} shows the growing backbones of the networks for the different decades of evolution. As time evolves, the backbone structure grows (number of nodes and edges increase) and becomes more intricate. The number of nodes, unique edges, edge mentions, number of clusters, and the average number of neighbors a node possesses in the growing backbone structures as shown in Fig. \ref{fig:backbone_evolution}, are summarized in Table \ref{table:table1}. Interestingly, the number of source-target pairs in the backbone structures-- indicated by the number of edges, is around $4\%$ for most years. However, their frequencies of engagement--  indicated by the number of edge mentions, grows steadily from $38\%$ (1970-1980) to $62\%$ (1970-2016). The average number of neighbors a node possesses, also increases from $1.537$ (1970-1980) to $1.817$ (1970-2016).

The backbones for the various decades contain between $8\%$ to $16\%$ of the total number of the nodes; the number of unique edges remains fairly constant around $4\%$ of the total number of edges in the network. Most importantly, the edge mentions (weights) grow from $38\%$ to $62\%$ of the total network, signifying very high frequency of engagement between a small number of source-target pairs. The terror hubs and vulnerable motifs were seen to have star structures more frequently than by chance. The average degree of a node in the backbone increased steadily as time evolved. 
The average clustering coefficient was always observed to be zero (indicating the absence of triangles or cyclicity) in the growing directed network \cite{fagiolo2007clustering,malliaros2013clustering}, as well as the evolving backbone. 

We analyzed the evolutionary structures of the hubs and motifs in Afghanistan, Colombia, India, Israel, Pakistan and the United Kingdom and found very interesting changes in the structures, especially in the Indian sub-continent. The backbone structures of the different countries significantly reflected the changes in the prominence of terrorist organizations in various decades.
We also observed that the US citizens, businessmen, and other organizations were often the common target nodes linking different closed knit communities of terrorist organizations from other countries. The observation that El Salvador appeared only in the backbones of 1970-1980 and 1970-1990 and not thereafter, conforming to the fact that Chapultepec peace accords were signed in 1992, is one of the many significant outcomes of the network backbone analysis. 
Our results for the range of provided parameters describe the evolution of terrorism in the above countries that emerge from the network analysis. The results are in no way a comment on the previous policies of the Governments of the countries considered.

The political and socioeconomic conditions along with the local circumstances of a region plays key role in framing anti-terrorism policies and elimination of terrorist ties. 
The inter-disciplinary approaches of network analysis that we have used in this paper, may provide supplementary knowledge and insight on the formation and spreading of terrorism, and thereby help the international security agencies in contending terrorism, as well as produce acumen for the policy makers and experts of international relations.

\subsection*{List of top 50 actor pairs}

\noindent
The list of top-50 actor pairs (source--target abbreviations) along with their weights (frequencies of interactions) that
appear in the backbone of the global terrorist network with 470 nodes and 427 unique edges is given in Table \ref{table:table3}.

\begin{table}[H]
\caption {List of top-50 actor pairs that appear in the backbone structure (1970-2016).}
\label{table:table3}
\vspace*{2mm}
\begin{small}
\centering
\begin{tabular}{|l|l|l|l|l|l|l|l|}
\hline
S. No. & Source          & Target    & Weight & S. No. & Source          & Target    & Weight \\ \hline
1      & Taliban\_AFG    & POL\_AFG  & 1854   & 26     & IRA\_GBR        & PCP\_NIRL & 296    \\ \hline
2      & ISIL\_IRQ       & PCP\_IRQ  & 1303   & 27     & NPA\_PHL        & POL\_PHL  & 293    \\ \hline
3      & Taliban\_AFG    & PCP\_AFG  & 1086   & 28     & Houthi\_YEM     & PCP\_YEM  & 292    \\ \hline
4      & Taliban\_AFG    & GOVG\_AFG & 810    & 29     & NPA\_PHL        & BUS\_PHL  & 291    \\ \hline
5      & BH\_NGA         & PCP\_NGA  & 740    & 30     & CPI-Maoist\_IND & GOVG\_IND & 278    \\ \hline
6      & SPSL\_PER       & PCP\_PER  & 733    & 31     & LTTE\_LKA       & PCP\_LKA  & 275    \\ \hline
7      & FMLN\_SLV       & UTL\_SLV  & 713    & 32     & IRA\_GBR        & BUS\_NIRL & 270    \\ \hline
8      & SPSL\_PER       & GOVG\_PER & 693    & 33     & TTP\_PAK        & PCP\_PAK  & 267    \\ \hline
9      & SPSL\_PER       & BUS\_PER  & 670    & 34     & FARC\_COL       & BUS\_COL  & 251    \\ \hline
10     & SPSL\_PER       & POL\_PER  & 552    & 35     & KWP(PKK)\_TUR   & PCP\_TUR  & 245    \\ \hline
11     & ETA\_ESP        & POL\_ESP  & 530    & 36     & FMLN\_SLV       & BUS\_SLV  & 243    \\ \hline
12     & SPSL\_PER       & UTL\_PER  & 506    & 37     & ELN\_COL        & UTL\_COL  & 230    \\ \hline
13     & ETA\_ESP        & BUS\_ESP  & 495    & 38     & IRA\_GBR        & BUS\_GBR  & 222    \\ \hline
14     & ISIL\_IRQ       & POL\_IRQ  & 451    & 39     & AQI\_IRQ        & PCP\_IRQ  & 221    \\ \hline
15     & CPI-Maoist\_IND & PCP\_IND  & 448    & 40     & Maoists\_IND    & PCP\_IND  & 219    \\ \hline
16     & IRA\_GBR        & POL\_NIRL & 433    & 41     & NPA\_PHL        & GOVG\_PHL & 218    \\ \hline
17     & FARC\_COL       & PCP\_COL  & 391    & 42     & BH\_NGA         & POL\_NGA  & 218    \\ \hline
18     & ASB\_SOM        & PCP\_SOM  & 387    & 43     & FARC\_COL       & UTL\_COL  & 215    \\ \hline
19     & KWP(PKK)\_TUR   & POL\_TUR  & 366    & 44     & FARC\_COL       & GOVG\_COL & 211    \\ \hline
20     & FE\_NGA         & PCP\_NGA  & 361    & 45     & LTTE\_LKA       & POL\_LKA  & 207    \\ \hline
21     & CPI-Maoist\_IND & POL\_IND  & 341    & 46     & FLNC\_FRA       & BUS\_FRA  & 204    \\ \hline
22     & FARC\_COL       & POL\_COL  & 333    & 47     & NPA\_PHL        & PCP\_PHL  & 200    \\ \hline
23     & FMLN\_SLV       & PCP\_SLV  & 330    & 48     & PE\_GBR         & PCP\_NIRL & 196    \\ \hline
24     & ASB\_SOM        & GOVG\_SOM & 324    & 49     & TTP\_PAK        & POL\_PAK  & 192    \\ \hline
25     & Maoists\_IND    & POL\_IND  & 315    & 50     & ETA\_ESP        & GOVG\_ESP & 190    \\ \hline
\end{tabular}
\end{small}
\end{table}
\section*{Acknowledgements}

The authors thank S. Biswas, A.S. Chakrabarti, B.K. Chakrabarti, A. Chatterjee, I. Ghosh, A. Krishnamachari, A.S. Patel, H.K. Pharasi, M.S. Santhanam, and P. Sen for useful comments and criticisms. The authors acknowledge the support by University of Potential Excellence-II grant (Project ID-47) of JNU, New Delhi, and the DST-PURSE grant given to JNU by the Department of Science and Technology, Government of India. K.S. acknowledges the University Grants Commission (Ministry of Human Resource Development, Govt. of India) for her senior research fellowship. 


\bibliographystyle{10}
\bibliography{GTD_Main}


%

\vskip 1in

\renewcommand{\thefigure}{S\arabic{figure}}
\renewcommand{\thetable}{S\arabic{table}}

\noindent
{\large \textbf{Supplementary information 
}}

\vskip 0.2in
\noindent


\section*{Data}
\begin{figure}[H]
\centering
\includegraphics[width=0.9\linewidth]{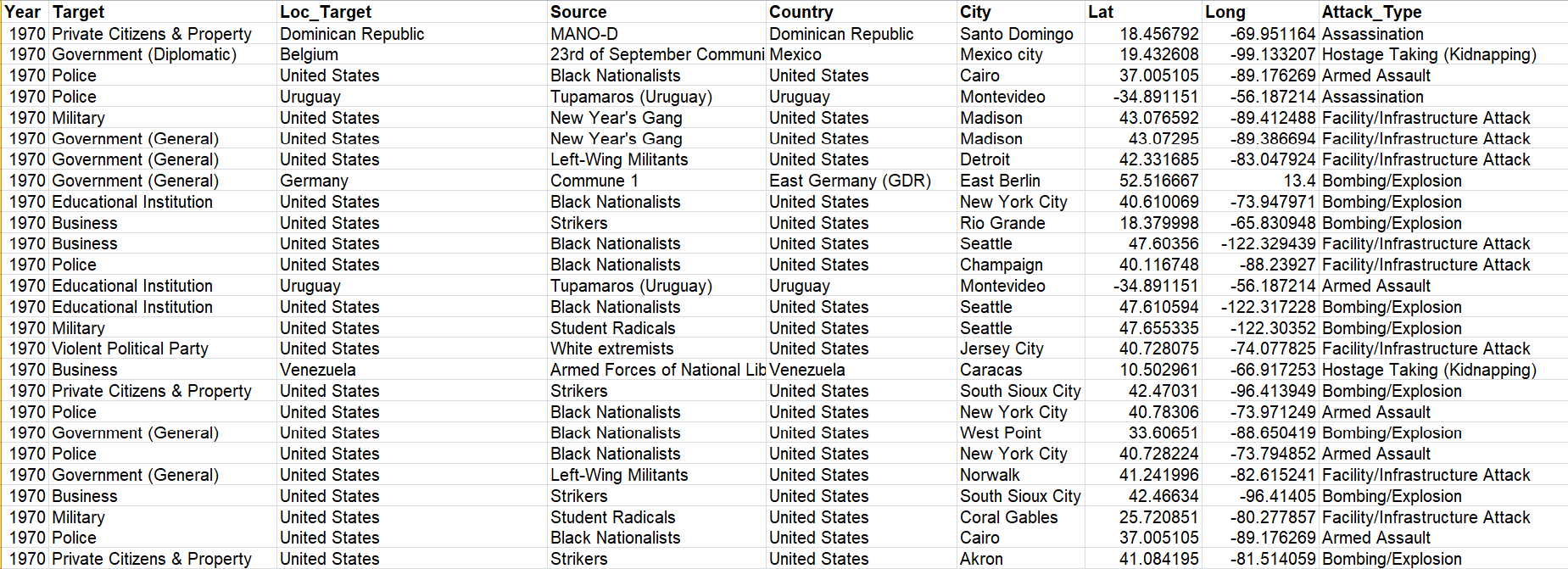}
\caption{Snapshot displaying few representative columns of the dataset.}
\label{fig:snip}
\end{figure}

The data source utilized for this quantitative analysis of terrorism is obtained from the Global Terrorism Database (GTD), maintained by the National Consortium for the Study of Terrorism and Responses to Terrorism (START) at the University of Maryland, United States. The database was built on unclassified source material publicly available in media, digital news archives, books, journals, and some legal documents. GTD contains 170350 terrorist events reported for a period of 46 years from 1970 to 2016. The events of 1993 are not present in the database as they were lost prior to START's compilation. The dataset includes 135 variables such as GTD Id, date of incident, incident location, incident information, attack information, target/victim information, perpetrator information, perpetrator statistics, claims of responsibility, weapon information, casualty information, consequences, kidnapping/hostage taking information, additional information, and source information. A snapshot of few variables is shown in Figure ~ \ref{fig:snip}.
\section*{Extended results: Figures and tables}
\section*{Terrorist attacks}
Different statistics and details of the terrorist attacks from 1970-2016, are shown below in Figure ~\ref{fig:yearwise}.
\begin{figure}[H]
\centering
\includegraphics[width=1\linewidth]{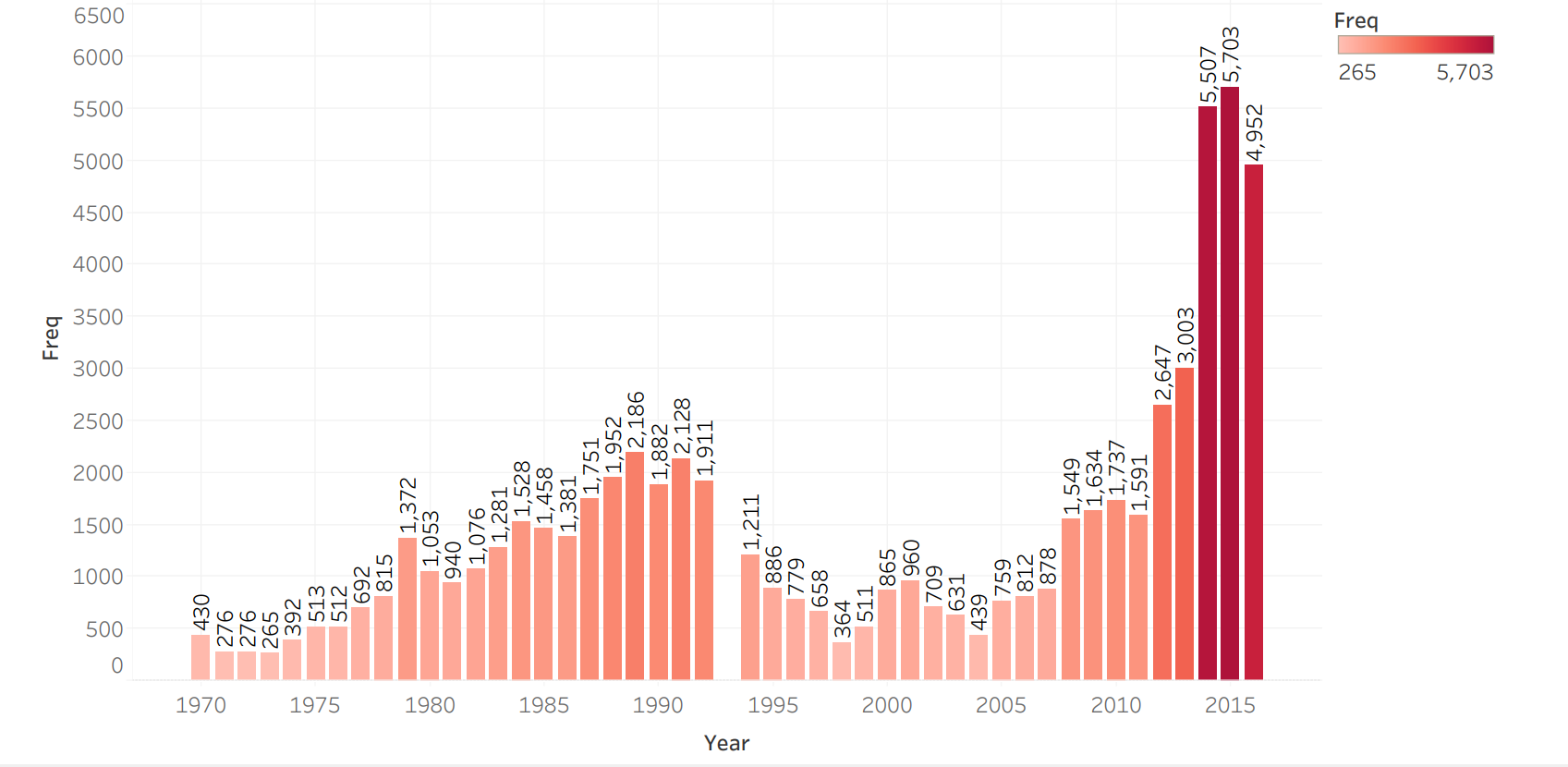}\llap{\parbox[b]{5.2in}{{\textbf{(a)}}\\\rule{0ex}{2.4in}}}
\includegraphics[width=0.6\linewidth]{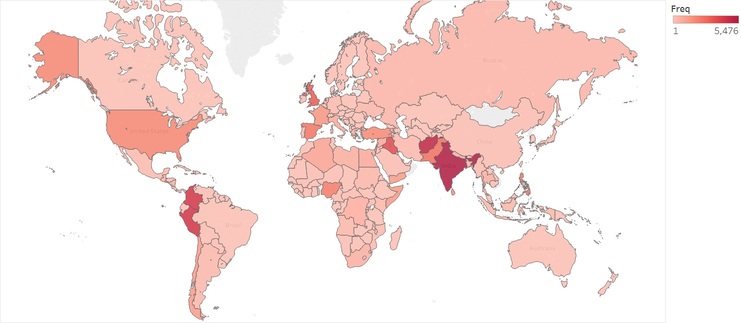}\llap{\parbox[b]{3.5in}{{ \textbf{(b)}}\\\rule{0ex}{1.4in}}}
\includegraphics[width=0.3\linewidth]{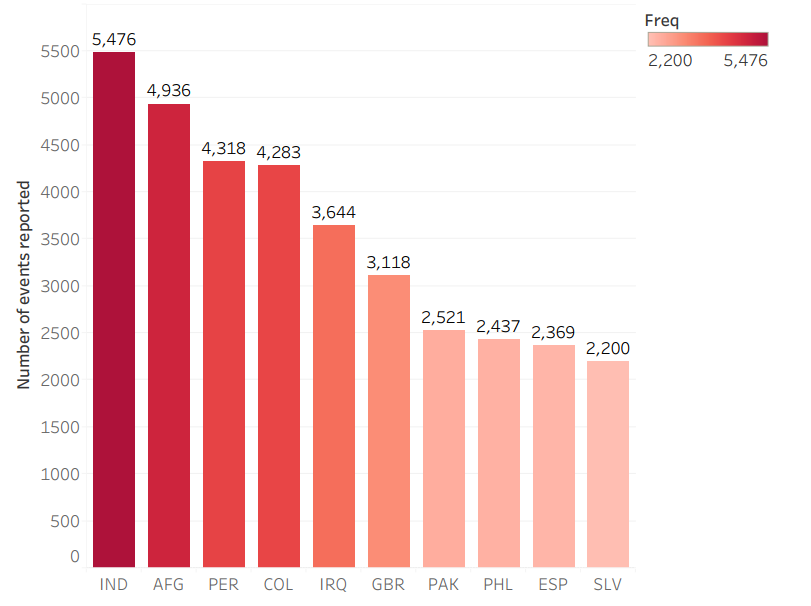}\llap{\parbox[b]{1.5in}{{ \textbf{(c)}}\\\rule{0ex}{1.4in}}}
\\\includegraphics[width=0.6\linewidth]{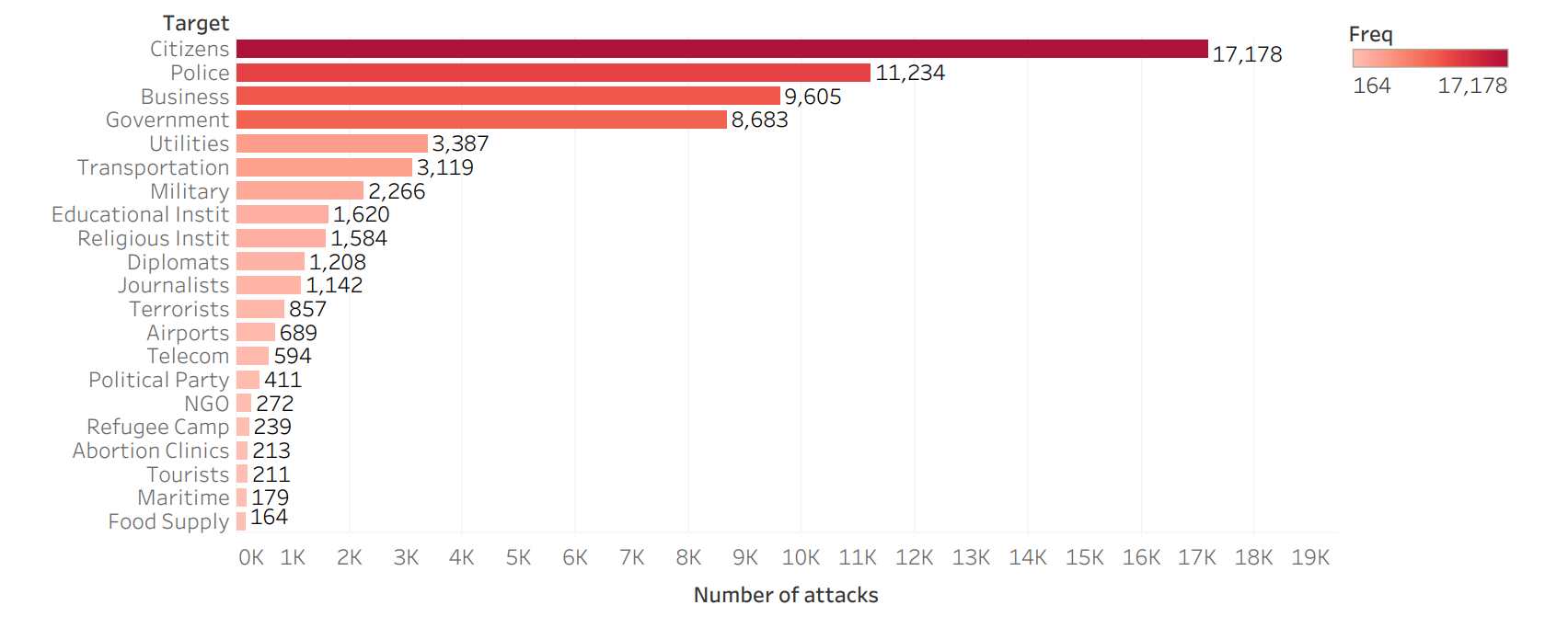}\llap{\parbox[b]{3.5in}{{ \textbf{(d)}}\\\rule{0ex}{1.4in}}}
\includegraphics[width=0.3\linewidth]{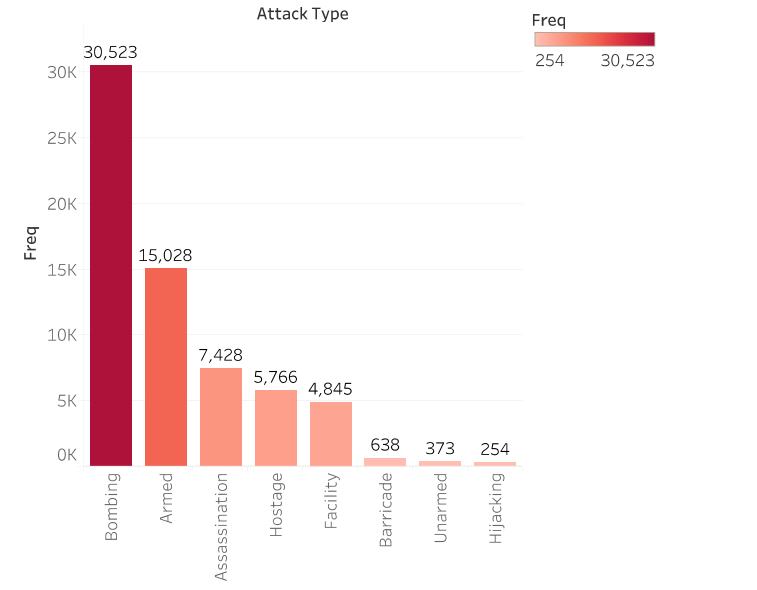}\llap{\parbox[b]{1.5in}{{ \textbf{(e)}}\\\rule{0ex}{1.4in}}}
\caption{(a) Year-wise count of terror activities. Note that no data was available for the year 1993. (b) Heatmap of the event count of global terrorism, showing the intensity and distribution. (c) Count of terror activities in the top-10 affected countries; note the plot depicts the raw figures for each country (not normalised by population or geographical area). (d) Target types-- most frequent targets of terrorists,  and (e) Attack types-- favorite \textit{modus operandi} for assaults. All results are for the period 1970-2016. The maps and international boundaries are generated using the proprietary software \textit{Tableau Desktop version 10.5.0}.}
\label{fig:yearwise}
\end{figure}

\subsection*{Fatalities vs. injured}
\begin{figure}[!h] 
\includegraphics[width=0.45\linewidth]{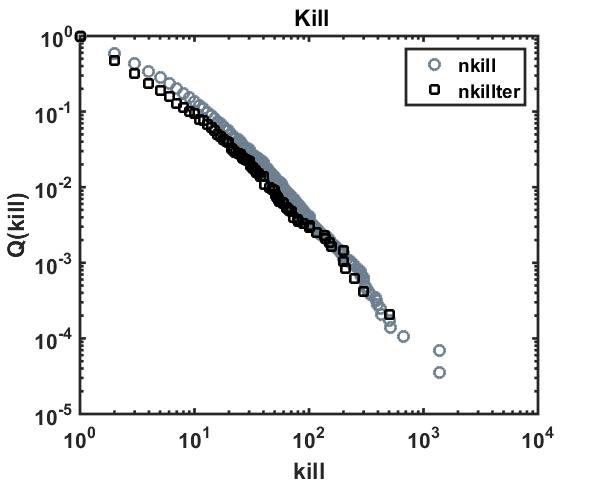}\llap{\parbox[b]{2.2in}{(\textbf{a})\\\rule{0ex}{1.8in}}}
\includegraphics[width=0.45\linewidth]{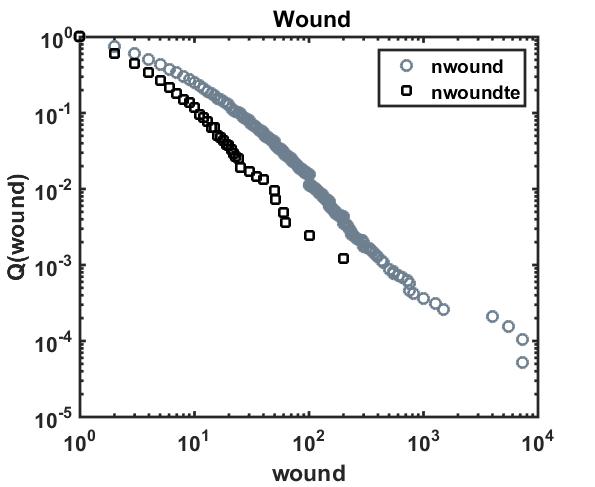}\llap{\parbox[b]{2.2in}{(\textbf{b})\\\rule{0ex}{1.8in}}}
\caption{Plots of the cumulative probability (CCDF) of (a) killed and (b) wounded for aggregated period over 1970-2016.}
\label{fig:kill}
\end{figure}

The impact of the terrorist attacks, as given by the number of persons killed or wounded, are shown in Figure ~\ref{fig:kill}. Interestingly, they have broad distributions. 

\subsection*{Evolution of Network and Giant Component}
\begin{figure}[h!]
\centering
\includegraphics[width=0.95\linewidth]{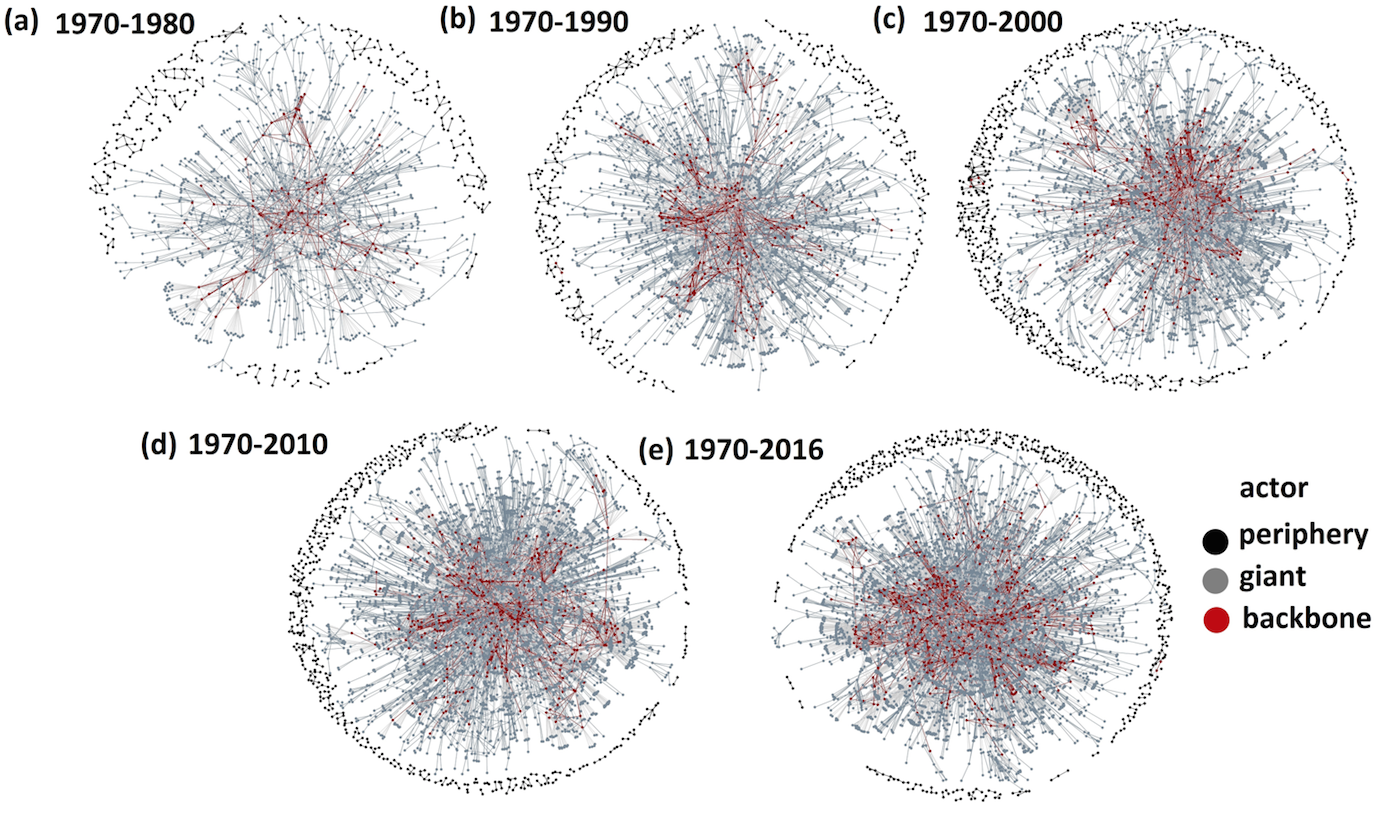}
\caption{Plots (a -- e) shows decade-wise evolution of the network, its giant component (in grey) as well as the backbone (in red). The network consists of 1459 nodes in 1970-1980, 2658 nodes in 1970-1990, 3832 nodes in 1970-2000, 4669 nodes in 1970-2010, and 5568 nodes in 1970-2016. The size of the giant component (shown in grey) and its percentage size with respect to the entire network is 1210 (83\%) in 1970-1980, 2346 (88\%) in 1970-1990, 3313 (86\%) in 1970-2000, 4220 (90\%) in 1970-2010, and 5148 (92\%) in 1970-2016. Interestingly, the backbone identified using the disparity filter and shown in red is primarily a part of the giant component. The peripheral nodes are colored black.}
\label{fig:giant_comp}
\end{figure}

\begin{figure}[H]
\centering
\includegraphics[width=0.23\linewidth]{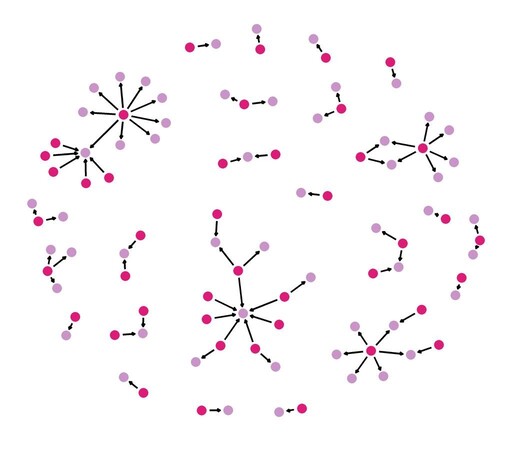}
\llap{\parbox[b]{2.0in}{\textbf{1970-1980}\\\rule{0ex}{0.5in}}}
\llap{\parbox[b]{0.8in}{\textbf{$\alpha_c=0.02$}\\\rule{0ex}{1.0in}}}
\includegraphics[width=0.23\linewidth]{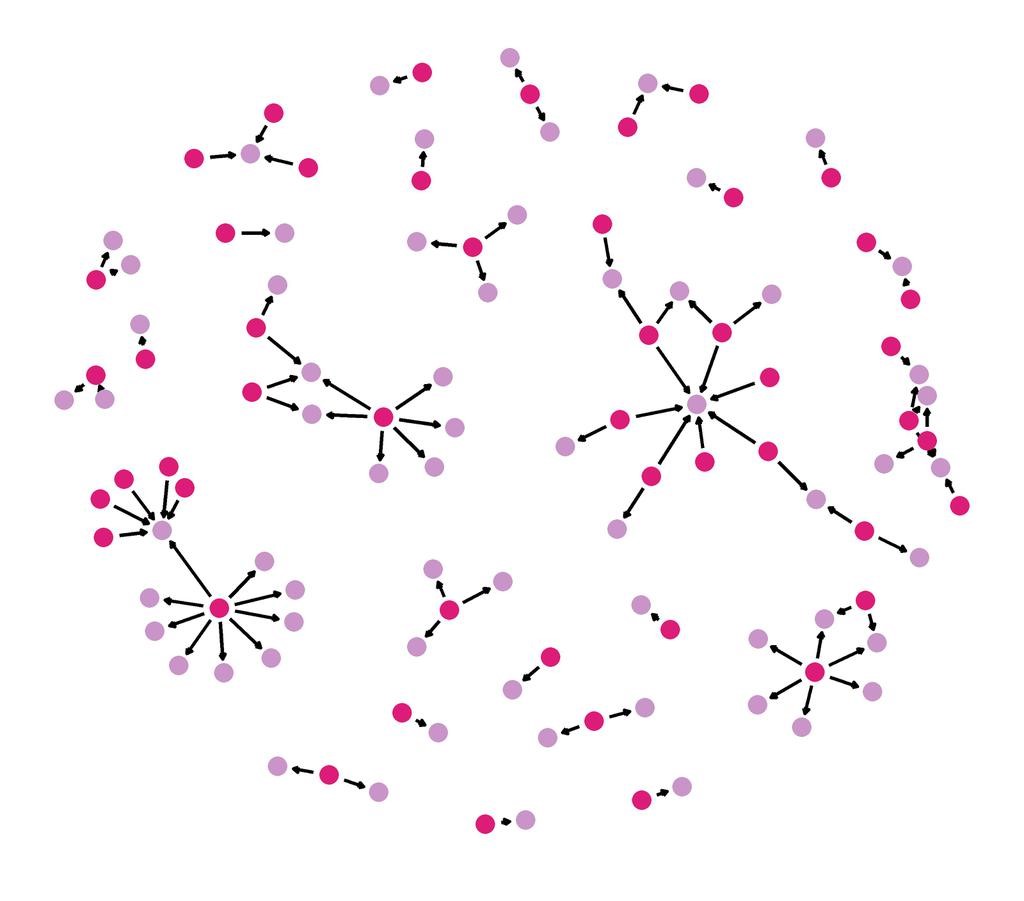}
\llap{\parbox[b]{0.8in}{\textbf{$\alpha_c=0.03$}\\\rule{0ex}{1.0in}}}
\includegraphics[width=0.23\linewidth]{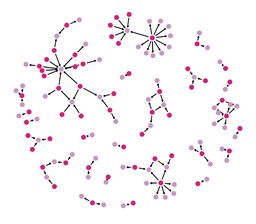}
\llap{\parbox[b]{0.8in}{\textbf{$\alpha_c=0.04$}\\\rule{0ex}{1.0in}}}
\includegraphics[width=0.23\linewidth]{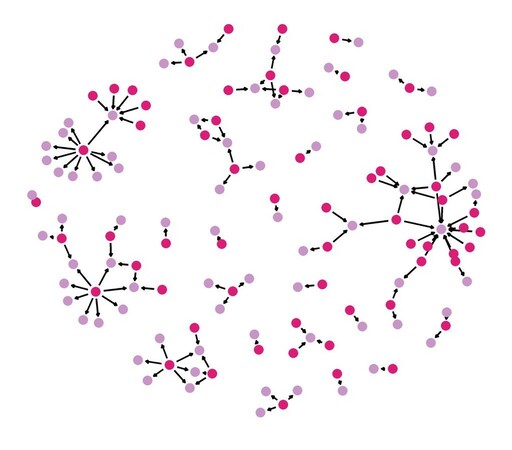}
\llap{\parbox[b]{0.8in}{\textbf{$\alpha_c=0.05$}\\\rule{0ex}{1.0in}}}\\

\includegraphics[width=0.23\linewidth]{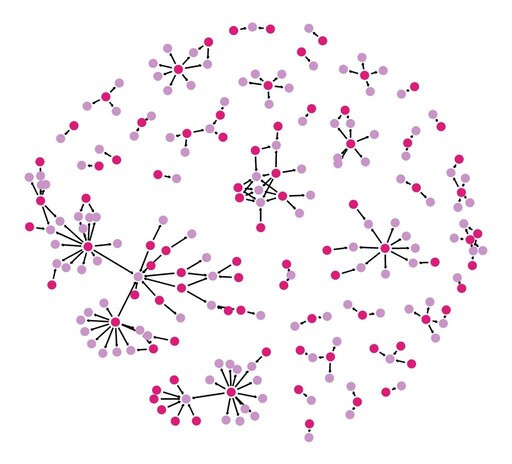}
\llap{\parbox[b]{2.1in}{\textbf{1970-1990}\\\rule{0ex}{0.5in}}}
\includegraphics[width=0.23\linewidth]{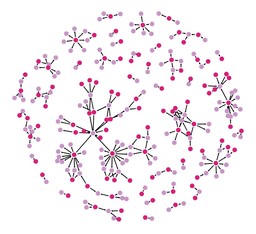}
\includegraphics[width=0.23\linewidth]{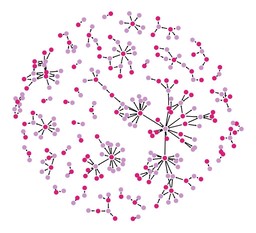}
\includegraphics[width=0.23\linewidth]{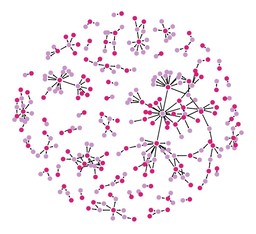}\\
\includegraphics[width=0.23\linewidth]{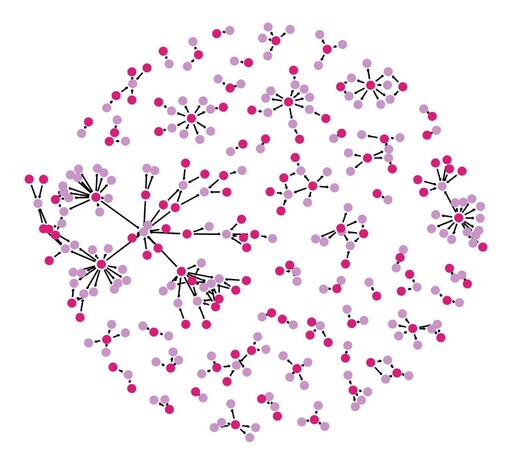}
\llap{\parbox[b]{2.1in}{\textbf{1970-2000}\\\rule{0ex}{0.5in}}}
\includegraphics[width=0.23\linewidth]{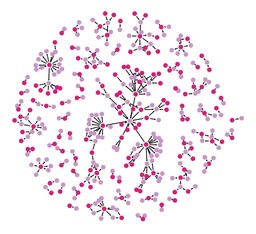}
\includegraphics[width=0.23\linewidth]{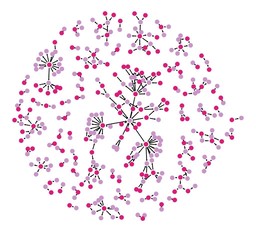}
\includegraphics[width=0.23\linewidth]{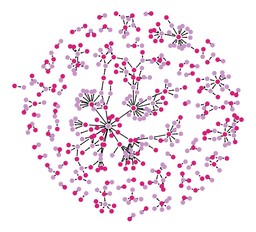}\\
\includegraphics[width=0.23\linewidth]{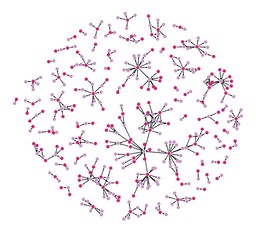}
\llap{\parbox[b]{2.1in}{\textbf{1970-2010}\\\rule{0ex}{0.5in}}}
\includegraphics[width=0.23\linewidth]{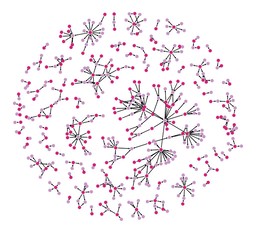}
\includegraphics[width=0.23\linewidth]{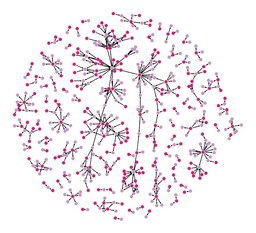}
\includegraphics[width=0.23\linewidth]{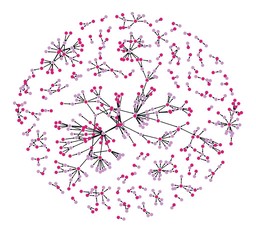}\\
\includegraphics[width=0.23\linewidth]{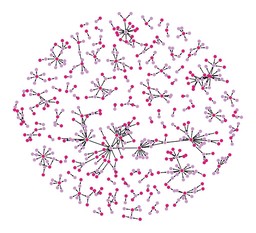}
\llap{\parbox[b]{2.1in}{\textbf{1970-2016}\\\rule{0ex}{0.5in}}}
\includegraphics[width=0.23\linewidth]{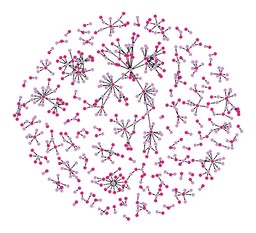}
\includegraphics[width=0.23\linewidth]{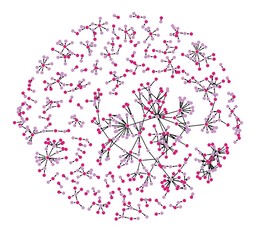}
\includegraphics[width=0.23\linewidth]{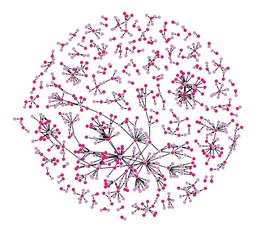}\\
\includegraphics[width=0.2\linewidth]{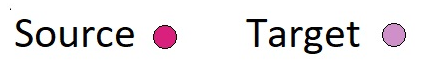}
\caption{Effect of $\alpha_c$ on the evolving backbones.  (\textit{From Left to Right}) Different values of $\alpha_c=0.02, 0.03, 0.04, 0.05$. (\textit{From Top to Bottom}) Different decades 1970-1980, 1970-1990, 1970-2000, 1970-2010 and 1970-2016, respectively.}
\label{fig:Backbone_alpha}
\end{figure}

Decade-wise evolution of the network and its giant component is shown in Figure \ref{fig:giant_comp}.
\newpage
The backbone structures are displayed in the Figure \ref{fig:Backbone_alpha}  as an effect of $\alpha_c$ on the evolving backbones

\newpage

\subsection*{Lists of names of sources, targets and actor pairs}

\noindent
The names of 190 sources and their abbreviations that appear in the backbone of the global terrorist network with 470 nodes are given in Table \ref{my-label}.


\begin{longtable}{|l|l|l|}
\caption{Names and abbreviations of Sources that appear in the backbone structure (1970-2016).}
\label{my-label}\\

\hline
\textbf{Sr. No.} & \textbf{Source Name}                                                                   & \textbf{Source Abbreviation} \\ \hline
1                & 23rd of September Communist League\_Mexico                                             & 23SepCL\_MEX                 \\ \hline
2                & 31 January People's Front (FP-31)\_Guatemala                                           & FP-31\_GTM                   \\ \hline
3                & Abu Sayyaf Group (ASG)\_Philippines                                                    & ASG\_PHL                     \\ \hline
4                & Action Directe\_France                                                                 & AcDir\_FRA                   \\ \hline
5                & African National Congress (South Africa)\_South Africa                                 & ANC\_ZAF                     \\ \hline
6                & Ajnad Misr\_Egypt                                                                      & AjM\_EGY                     \\
\end{longtable}

\noindent
The names of 280 targets and their abbreviations that appear in the backbone of the global terrorist network with 470 nodes are given in Table \ref{my-label1}.

\begin{longtable}{|l|l|l|}

\caption{Names and abbreviations of Targets that appear in the backbone structure (1970-2016)}
\label{my-label1}\\


\hline
\textbf{Sr. No.} & \textbf{Target Name}                                           & \textbf{Target} \\ \hline
1                & Government (General)\_Afghanistan                              & GOVG\_AFG       \\ \hline
2                & Police\_Afghanistan                                            & POL\_AFG        \\ \hline
3                & Private Citizens \& Property\_Afghanistan                      & PCP\_AFG        \\ \hline
4                & Police\_Algeria                                                & POL\_DZA        \\ \hline
5                & Business\_United States                                        & BUS\_USA        \\ \hline
6                & Private Citizens \& Property\_Burundi                          & PCP\_BDI        \\
\end{longtable}

\end{document}